\documentclass[pra,twocolumn,aps,amssymb,showpacs,superscriptaddress]{revtex4-1}

\usepackage{epsfig}
\usepackage{graphicx}
\usepackage{amsmath} 
\usepackage{amssymb}
\usepackage{enumerate}
\usepackage{hyperref}
\usepackage[normalem]{ulem}
\usepackage{cancel}

\usepackage{color}
\definecolor{rosso}{rgb}{1,0,0}
\definecolor{verde}{rgb}{0,1,0}
\definecolor{blue}{rgb}{0,0,1}
\definecolor{verdescuro}{rgb}{0,0.5,0.5}
\definecolor{rossoscuro}{rgb}{0.7,0.3,0}
\definecolor{bluscuro}{rgb}{0.3,0,0.7}
\definecolor{magenta}{rgb}{1,0,1}

\begin{document}

\title{Recent advances in the theory of the BCS-BEC crossover for fermionic superfluidity}

\author{Giancarlo Calvanese Strinati}
\email{giancarlo.strinati@unicam.it}
\affiliation{School of Science and Technology, Physics Division, Universit\`{a} di Camerino, 62032 Camerino (MC), Italy}
\affiliation{CNR-INO, Istituto Nazionale di Ottica, Sede di Firenze, 50125 (FI), Italy}


\begin{abstract}
The BCS–BEC crossover realized experimentally with ultra-cold Fermi gases may be considered as one of the important scientific achievements occurred during the last several years.
The flexibility for operating on these systems on the experimental side and the full control of the relevant system degrees of freedom on the theoretical side make quite stringent at a fundamental level the comparison 
between the experimental data and the corresponding theoretical calculations. 
Here, we briefly survey recent theoretical advances resting on a diagrammatic approach at equilibrium that improves in a systematic way on the widely used $t$-matrix approach, yielding a quite good comparison between theory and
experiments for several physical quantities of interest.
It is proposed that the physical phenomena underlying this theoretical approach may also be relevant to the superconducting phase of condensed-matter materials which cannot be described by the standard BCS theory.
\vspace{0.1cm}

\end{abstract}

\maketitle

The marked similarity between bosonic and fermionic superfluidity has been recognized for some time \cite{Tilley-Tilley-1986}, although these phenomena were originally discovered experimentally in quite different physical systems 
(namely, $^{4}\mathrm{He}$ and superconductors, respectively).
In ultimate analysis, the similarity stems from these different physical systems sharing the basic feature of having the same kind of spontaneously broken symmetry.

More recently, a closer connection between bosonic and fermionic superfluidity has been experimentally realized with ultra-cold Fermi gases, for which by tuning the inter-particle interaction \emph{a single system\/} evolves continuously,
from a BCS state where pairs of (opposite spin) fermions are described by Fermi statistics, to a BEC state where two-fermion dimers are described by Bose statistics. 
Here, BCS refers to the fundamental theory of superconductors by the Bardeen, Cooper, and Schrieffer \cite{BCS-1957}, while BEC refers to the Bose-Einstein condensation (see, e.g., Ref.~\cite{PS-2003}).
This continuous evolution has been dubbed the \emph{BCS–BEC crossover\/}, since it corresponds to a situation when the system goes from one phase to another as a certain parameter is changed 
without encountering a phase transition in between.
For ultra-cold Fermi gases this parameter is represented by the inter-particle coupling $(k_{F} a_{F})^{-1}$, where $k_{F} = \left(3 \pi^{2} n \right)^{1/3}$ is the Fermi wave vector for particle density $n$ and $a_{F}$ is the scattering
length of the two-fermion problem in vacuum.
It is the value of $a_{F}$ that varies by spanning a molecular Fano-Feshbach resonance \cite{Fano-1961,Feshbach1962} (typically, a broad resonance of $^{6}\mathrm{Li}$ has conveniently been utilized \cite{Grimm-2010}).
A concise summary of how the BCS–BEC crossover has been experimentally realized can be found in Ref.~\cite{Physics-Reports-2018}.

In this last reference, a summary is also given about theoretical approaches to the BCS–BEC crossover based on diagrammatic approximations.
These approaches are relevant because, to evolve from the weak-coupling (BCS) limit to the strong-coupling (BEC) limit, consideration of pairing fluctuations over and above the original mean-field approach of Ref.~\cite{BCS-1957} is required
especially at finite temperature.
This crucial feature was first pointed out by Nozi\`{e}res and Schmitt-Rink in their pioneering work on the $t$\emph{-matrix approach\/} for the BCS–BEC crossover in the normal phase \cite{NSR-1985}, 
an approach that was later extended in several ways by considering all possible different shades of self-consistency in the fermionic lines entering the $t$-matrix (a comparative study in this respect is given in Ref.~\cite{PPS-2019}).
The $t$-matrix approach (or, else, the ladder approximation) was actually introduced by Galitskii to deal with a dilute Fermi gas with repulsive inter-particle interaction, for which $a_{F} > 0$ and $k_{F} a_{F} \ll 1$ \cite{Galitskii-1958}.
Just after, to deal with the phenomenon of fermionic superfluidity, the ladder approximation was extended by Gorkov and Melik-Barkhudarov to the case of a short-range attractive inter-particle interaction for which $a_{F} < 0$ \cite{GMB-1961}, 
but only in the limit $k_{F} |a_{F}| \ll 1$ that corresponds to the far BCS side of the BCS–BEC crossover.
A self-consistent version of the $t$-matrix approach was also considered in Ref.~\cite{Haussmann-2007} for the superfluid phase below the transition temperature $T_{c}$.
\begin{figure}[h]
\includegraphics[width=6.5cm,angle=0]{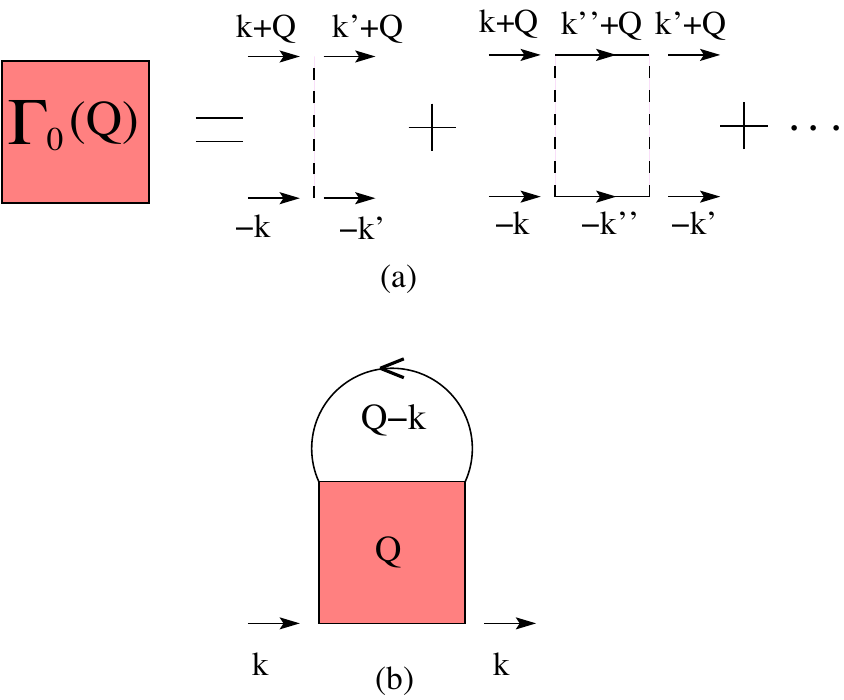}
\caption{Diagrammatic representation of (a) the pair propagator $\Gamma_{0}$ for opposite-spin fermions and (b) the fermionic single-particle self-energy obtained from $\Gamma_{0}$ within the $t$-matrix approach. 
              Solid and dashed lines stand for the fermionic single-particle propagator and the interaction potential, and $Q (k)$ is a bosonic (fermionic) four-vector. 
              [\emph{Source}: Reproduced from Ref.~\cite{PPPS-2018}.]} 
\label{Figure-1}
\end{figure} 
For later reference, \color{red}Fig.~\ref{Figure-1} \color{black}shows the diagrammatic representation of the single-particle fermionic self-energy in terms of the pair ladder propagator associated with the $t$-matrix approach,
where different variants of this approach dress differently the single-particle lines therein (as discussed in detail in Ref.~\cite{PPS-2019}).

\begin{figure}[t]
\includegraphics[width=6.3cm,angle=0]{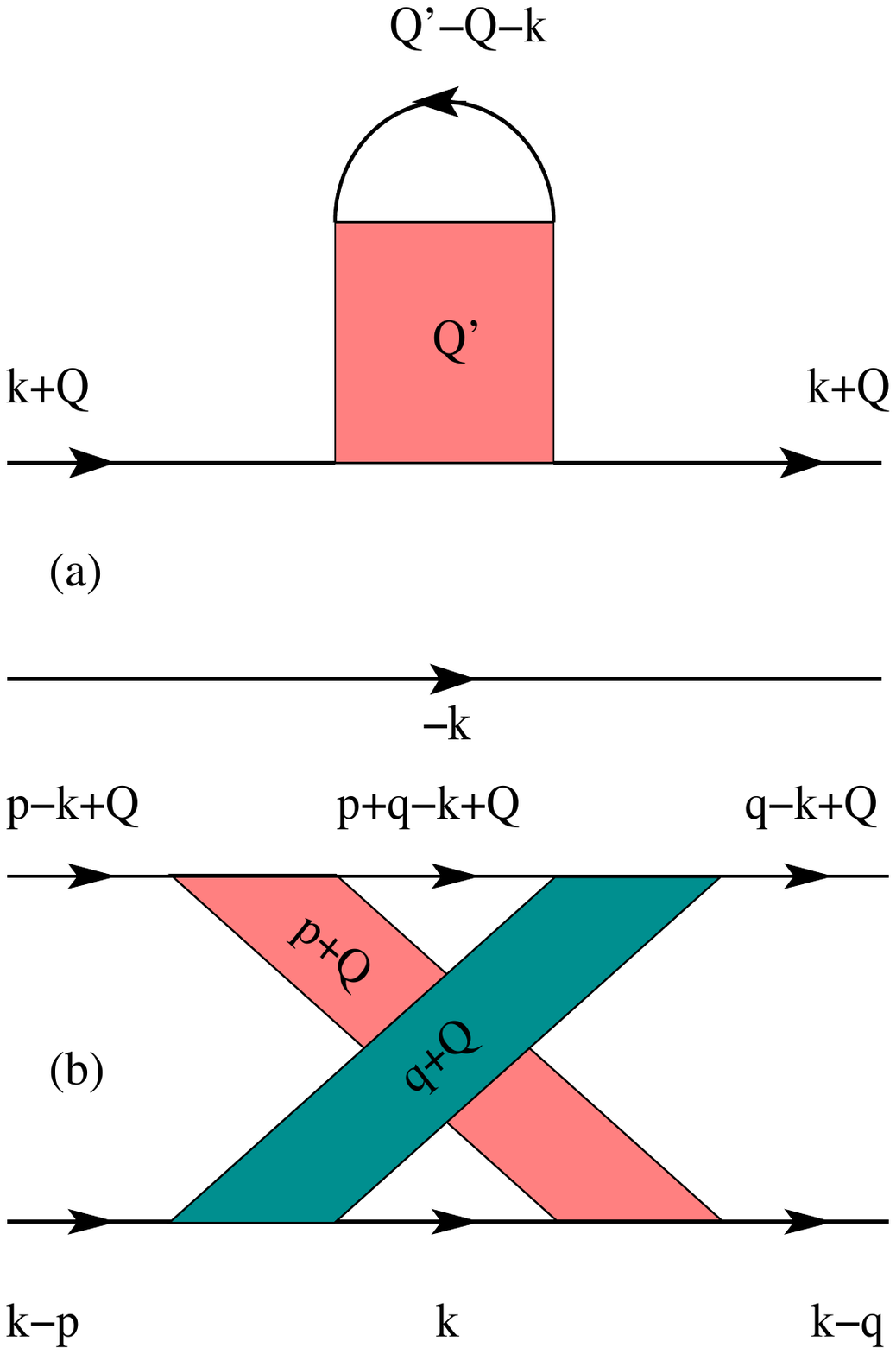}
\caption{Diagrammatic representation of (a) the Popov bosonic-like self-energy obtained by dressing the upper fermionic line in the particle-particle rung (an analogous dressing occurs for the lower fermionic line), 
               and (b) the GMB bosonic-like self-energy (where upper and lower fermionic lines correspond to opposite spins).
               Both (a) and (b) diagrams represent bosonic-like self-energy insertions to the pair propagator $\Gamma_{0}$ of Fig.~\ref{Figure-1}. 
               [\emph{Source}: Reproduced from Ref.~\cite{PPPS-2018}.]} 
\label{Figure-2}
\end{figure}

Quite generally, one of the main advantages for adopting a diagrammatic approach is that this approach can be dealt with in a ``modular way'', to the extent that a given diagrammatic approximation can be suitably improved by considering additional diagrammatic contributions that are relevant the physical problem one is considering.
Specifically, for the $t$-matrix approach to a dilute Fermi gas with a quite small Fermi surface such that only particle-particle rungs are retained to begin with as shown in \color{red}Fig.~\ref{Figure-1}\color{black}, an improvement 
over this approximation corresponds to including also particle-hole rungs in an appropriate fashion.
Although this inclusion might \emph{a priori\/} be considered to produce only minor corrections to physical quantities, Gorkov and Melik-Barkhudarov (GMB) showed that this is not the case, because it actually leads to a sizable 
reduction of the values of $T_{c}$ and of the BCS gap parameter $\Delta_{0}$ at zero temperature \cite{GMB-1961}.
This can be seen as follows.
Consider the expression of $T_{c}$ as obtained by the BCS theory \cite{BCS-1957}
\begin{equation}
k_{B} T_{c} = \frac{8 e^{\gamma} E_{F}}{\pi e^{2}} \exp\{\pi/(2 k_{F} a_{F})\} \, ,
\label{Tc-BCS}
\end{equation} 
where $k_{B}$ is the Boltzmann constant, $E_{F} = k_{F}^{2}/(2m)$ the Fermi energy ($m$ being the fermion mass), and $\gamma$ the Euler constant (with $e^{\gamma} \simeq 1.781$).
[We set $\hbar = 1$ throughout.]
Owing to the exponential dependence on coupling of this expression, if additional terms in the small parameter $k_{F} a_{F}$ are introduced in the exponent such that $(k_{F} a_{F})^{-1} \rightarrow (k_{F} a_{F})^{-1} + b + c \, (k_{F} a_{F}) + \cdots$ 
with $b$ and $c$ constants, the constant $b$ modifies the BCS pre-factor in Eq.~(\ref{Tc-BCS}) by a finite amount even in the (extreme) weak-coupling limit when $k_{F} a_{F} \rightarrow 0^{-}$.
To obtain the value of the constant $b$, GMB considered a correction to the BCS instability that occurs when $T_{c}$ is approached from the normal phase.
This instability was obtained in terms of ladder diagrams of \color{red}Fig.~\ref{Figure-1}(a)\color{black} (which corresponds to the so-called Thouless criterion \cite{Thouless-1960}), with additional contributions associated
with the particle-hole rungs mentioned above (to be discussed in more detail in \color{red}Fig.~\ref{Figure-2}(b) \color{black}below).
The end result of the GMB calculation for $T_{c}$ was a reduction of the expression (\ref{Tc-BCS}) for $T_{c}$ by the factor $(4e)^{1/3} \simeq 2.2$.
A similar reduction was also obtained by GMB for $\Delta_{0}$, such that the BCS value $3.52$ of the coupling ratio $2 \Delta_{0}/k_{B}T_{c}$ is not modified.

The original GMB correction to the BCS theory \cite{GMB-1961} addressed only the values of $T_{c}$ and $\Delta_{0}$ in the far BCS side of the BCS–BEC crossover where $(k_{F} a_{F})^{-1} \ll -1$.
Extension of the GMB correction to the whole BCS-BEC crossover would require one to modify the additional diagrammatic contribution (with respect to BCS theory) that was originally considered in Ref.~\cite{GMB-1961}, 
so as to include the full dependence on wave vector and frequency of the pair propagator $\Gamma_{0}$ of \color{red}Fig.~\ref{Figure-1}(a)\color{black}.
This extension was first considered for the normal phase above $T_{c}$ in Ref.~\cite{PPPS-2018}, and later adapted to the superfluid phase below $T_{c}$ in Ref.~\cite{PPS-2018}.
In particular, in Ref.~\cite{PPPS-2018} the diagram shown in \color{red}Fig.~\ref{Figure-2}(b) \color{black} was interpreted as being a bosonic-like self-energy for the pair propagator $\Gamma_{0}$, where the full wave-vector and frequency 
dependence of the two $\Gamma_{0}$ entering this diagram was retained (although in the original GMB correction of Ref.~\cite{GMB-1961} only the constant BCS result $\Gamma_{0} \simeq -4 \pi a_{F}/m$ was utilized).
In addition, in Ref.~\cite{PPPS-2018} it was found it necessary to include also the so-called Popov contribution shown in \color{red}Fig.~\ref{Figure-2}(a)\color{black}, that was introduced in Ref.~\cite{PS-2005} to account for the residual interaction between Cooper pairs in the BCS side and composite bosons in the BEC side of the crossover. 
The Popov contribution also acts to eliminate a spurious factor $e^{-1/3}$ obtained by the (bare) $t$-matrix approach for the expression of the critical temperature in the BCS limit.
[In superfluid phase, a corresponding GMB ``anomalous'' bosonic-like self-energy for the pair propagator need also be included \cite{PPS-2018}.]
Taken together, the Popov plus GMB contributions of \color{red}Fig.~\ref{Figure-2} \color{black} on top of the $t$-matrix approach are referred to as the \emph{extended GMB theory\/}.
We shall argue that this theory represents a valuable extension of the standard $t$-matrix approach, insofar as it is able to considerably improve the comparison of the numerical outcomes of the theory with the experimental data 
for important physical quantities
(like the coupling dependence of the low-temperature gap parameter $\Delta$ and of the critical temperature $T_{c}$, as well as  the temperature dependence of the superfluid density $\rho_{\mathrm{s}}$ at unitarity) 
that can accurately be measured in ultra-cold Fermi gases (see below).

\begin{figure}[t]
\includegraphics[width=8.5cm,angle=0]{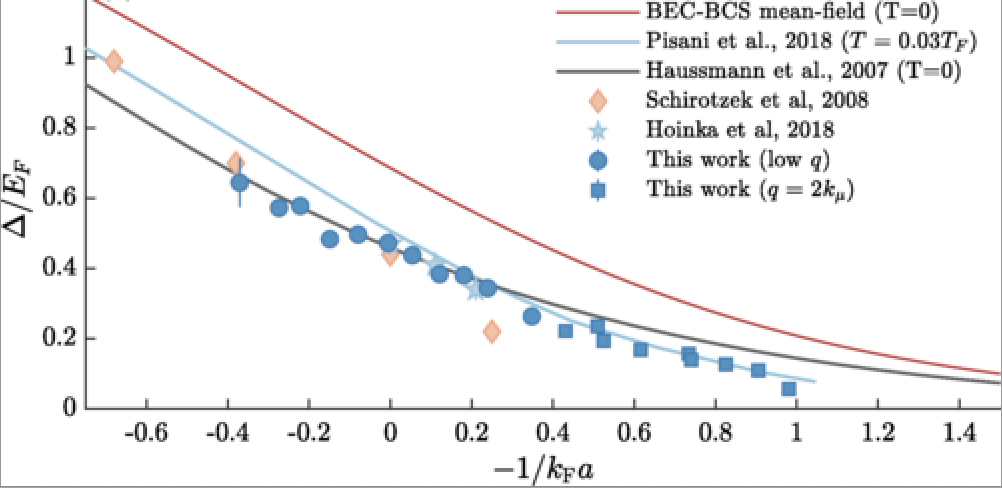}
\caption{Measurements of the low-temperature pairing gap $\Delta$ (in units of the Fermi energy $E_{F}$) obtained in Ref.~\cite{Moritz-2022} across the BCS-BEC crossover for a balanced spin mixture 
             of an ultra-cold gas of $^{6}\mathrm{Li}$ atoms (note the sign change for the inter-particle coupling in the horizontal axis with respect to our previous definition). 
             Comparison with three theoretical results is also reported (see the text for further details).
             Here, Pisani et al. 2018 refers to Ref.~\cite{PPS-2018}, Haussmann et al. 2007 to Ref.~\cite{Haussmann-2007}, Schirotzek et al. 2008 to Ref.~\cite{Schirotzek-2008}, and Hoinka et al. 2018 to Ref.~\cite{Hoinka-2018}. 
             [\emph{Source}: Reproduced from Ref.~\cite{Moritz-2022}.]} 
\label{Figure-3}
\end{figure} 

We begin by considering the coupling dependence of the gap parameter (or pairing gap) $\Delta$ at low temperature. 
This quantity was recently measured in Ref.~\cite{Moritz-2022}, where Bragg spectroscopy was used to obtain the momentum-resolved low-energy excitation spectrum of a balanced spin mixture of an ultra-cold gas of $^{6}\mathrm{Li}$ atoms,
with the pairing gap $\Delta$ being determined by fits to the excitation spectrum.
\color{red}Figure~\ref{Figure-3} \color{black} shows the coupling dependence of $\Delta$ reproduced from Fig.~4 of Ref.~\cite{Moritz-2022}, where the experimental data (collected also from other sources) are compared with the theoretical
results obtained in the superfluid phase by the self-consistent $t$-matrix approach of Ref.~\cite{Haussmann-2007} and by the extended GMB approach of Ref.~\cite{PPS-2018} (in addition to the standard mean-field results for which no paring
fluctuation is included).
Note, in particular, the good agreement between the experimental data and the theoretical results from Ref.~\cite{PPS-2018} especially on the BCS side of unitarity (while on the BEC side of unitarity comparison with the theoretical results from Ref.~\cite{Haussmann-2007} looks somewhat better).

\begin{figure}[t]
\includegraphics[width=8.5cm,angle=0]{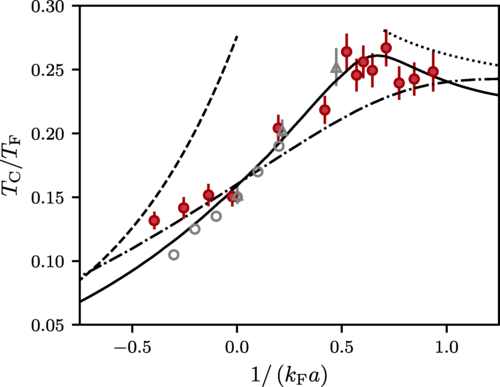}
\caption{Measurements of the critical temperature $T_{c}$ (in units of the Fermi temperature $T_{F}$) for an ultra-cold Fermi gas spanning the BEC-BCS crossover obtained in Ref.~\cite{Koehl-2023} are compared with the results of
             theoretical calculations (see the text for further details). 
             [\emph{Source}: Reproduced from Ref.~\cite{Koehl-2023}.]} 
\label{Figure-4}
\end{figure}

Even more recently, accurate measurements of the critical temperature $T_{c}$ across the BEC-BCS crossover (again for a balanced spin mixture of an ultra-cold gas of $^{6}\mathrm{Li}$ atoms) were reported in Ref.~\cite{Koehl-2023}, 
where a pioneering application of an artificial neural network was utilized to determine the phase diagram of strongly correlated fermions in the BCS-BEC crossover.
The corresponding results are reported in \color{red}Fig.~\ref{Figure-4}\color{black}.
Here, the experimental data from Ref.~\cite{Koehl-2023} show a steady increase of $T_{c}$ from the BCS side up to inter-particle coupling of approximately $0.5$, after which $T_{c}$ levels off and stays approximately constant or, possibly, 
decreases weakly for larger couplings.
The experimental data are also compared both with the theoretical results for $T_{c}$ obtained by the fully self-consistent $t$-matrix approach of Ref.~\cite{Haussmann-2007} coming from the superfluid phase (dashed-dotted line) 
which show a monotonic increase of $T_{c}$ with no maximum present, and by the extended GMB approach of Ref.~\cite{PPPS-2018} coming from the normal phase (full line) which instead leads to a very good agreement with the experimental data.
Note, in particular, how the extended GMB results of Ref.~\cite{PPPS-2018} are able to capture both the position (that occurs in the BEC side of unitarity) and the value of the maximum of $T_{c}$.
The outcomes of quantum Monte-Carlo calculations also reported in this figure appear to give further support to the results of the extended GMB results of Ref.~\cite{PPPS-2018}.

\begin{figure}[h]
\includegraphics[width=8.9cm,angle=0]{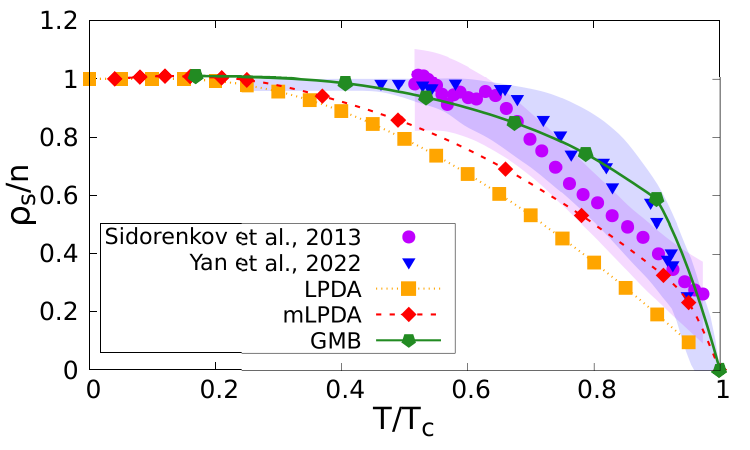}
\caption{Measurements of the temperature dependence of the superfluid density $\rho_{s}$ for a Fermi gas at unitarity, obtained in Ref.~\cite{Grimm-2013} (Sidorenkov {\emph et al.\/}) and Ref.~\cite{Zwierlein-2022} (Yan {\emph et al.\/}),
             are compared with the results of three theoretical calculations (see the text for further details). 
             In all cases, the temperature is in units of the corresponding value of the critical temperature $T_{c}$. 
             [\emph{Source}: Reproduced from Ref.~\cite{PPS-2023}.]}  
\label{Figure-5}
\end{figure}

An additional physical quantity of special importance for superfluid systems is the superfluid density $\rho_{s}$, whose temperature dependence in the homogeneous case decreases from the value of the particle density $n$ at zero temperature
down to zero at the critical temperature.
This quantity was measured for an ultra-cold gas of $^{6}\mathrm{Li}$ atoms at unitarity in both Refs.~\cite{Grimm-2013} and \cite{Zwierlein-2022}.
The corresponding experimental results are reported in \color{red}Fig.~\ref{Figure-5}\color{black},
where a comparison is also shown with the outcomes of the mean-field calculation (indicated here as LPDA), the $t$-matrix approach of Ref.~\cite{PPS-2004} (indicated here as mLPDA), 
and the extended GMB approach of Ref.~\cite{PPPS-2018} (indicated here as GMB).
Even as far as the temperature dependence $\rho_{s}$ is concerned, the GMB approach appears to provide the best comparison with the available experimental data.

Nonetheless, one should mention that in all versions of the $t$-matrix approach, either non-self consistent \cite{PPS-2004} or fully self-consistent \cite{Haussmann-2007}, close enough to $T_{c}$ the gap parameter $\Delta$ 
turns out to be a multivalued function of temperature, with a re-entrant behavior reminiscent of a first-order transition
(cf., e.g., Fig.~8 from Ref.~\cite{Haussmann-2007} and Fig.~7 from Ref.~\cite{PPS-2018}).
Overcoming this unwanted feature should possibly require the inclusion of additional (although not yet identified) diagrammatic contributions beyond the ladder structure of the $t$-matrix (even beyond the Popov an GMB contributions). 
It turns out that the magnitude of this re-entrant behavior for $\Delta$ gets amplified from the BCS to the BEC sides of the crossover, and eventually decreases in the extreme BEC limit.
Since this feature is unavoidably shared by the extended GMB approach of Ref.~\cite{PPS-2018}, close enough to $T_{c}$ the numerical results obtained by this approach may not be fully reliable, although in practice to a different extent 
depending on the physical quantity at hand.
For instance, the temperature dependence of the condensate fraction $n_{0}$ calculated by the extended GMB approach appears not to be too much influenced by the re-entrance behavior of the gap parameter $\Delta$, 
as it was shown in Ref.~\cite{PPS-2022}.
It is for this reason that in \color{red}Fig.~\ref{Figure-5} \color{black} for caution the last point calculated in terms of the extended GMB approach stops at $T/T_{c}=0.95$.

\begin{figure}[t]
\includegraphics[width=8.8cm,angle=0]{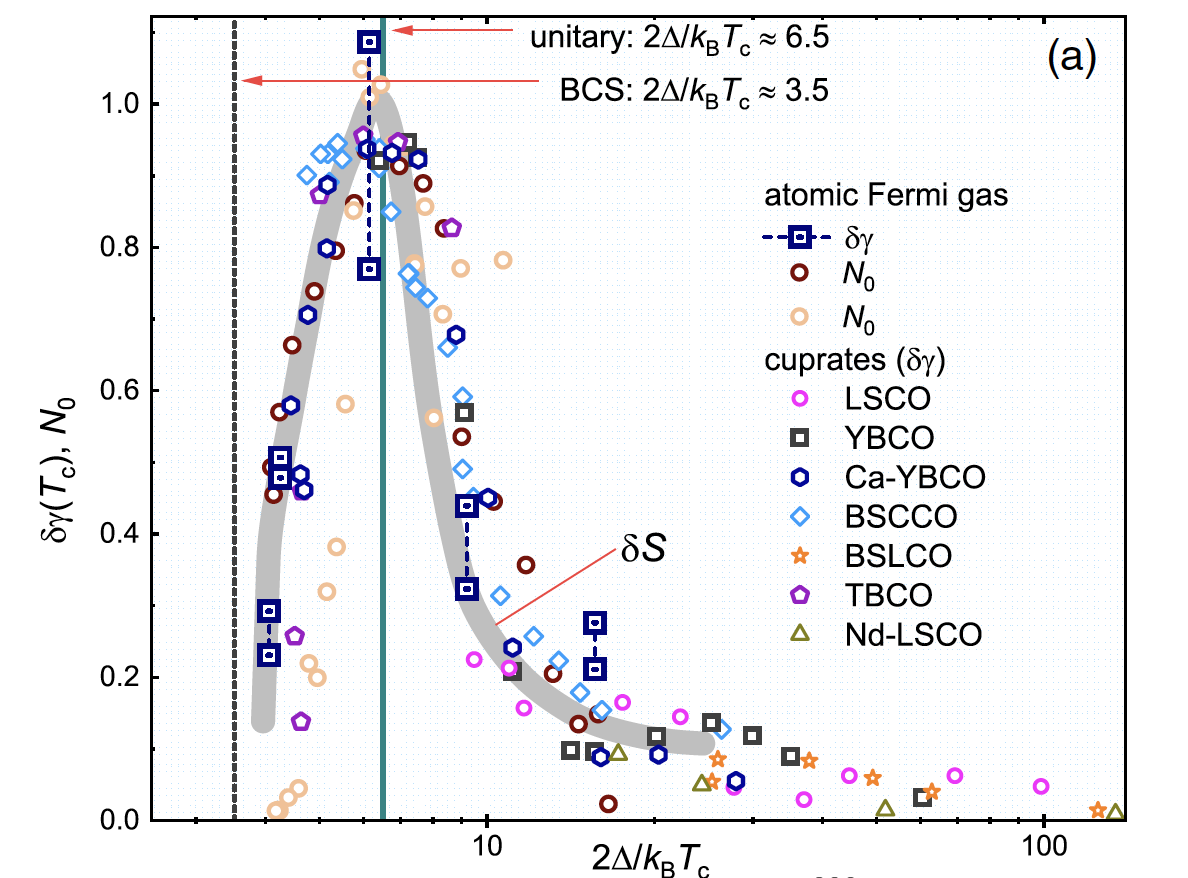}
\caption{The height of the jump $\delta \gamma (T_{c})$ in the electronic contribution to the specific heat at $T_{c}$ is shown vs the coupling ratio $2 \Delta/k_{B} T_{c}$ for various high-temperature cuprate superconductors.
             Here, $\Delta$ refers to the magnitude of the pairing gap at low temperature, such that it does not deviates considerably from its zero-temperature value $\Delta_{0}$. 
             [\emph{Source}: Reproduced from Ref.~\cite{Harrison-Chan-2022}.]} 
\label{Figure-6}
\end{figure} 

Thus far, ultra-cold Fermi gases are the physical systems for which the BCS-BEC crossover has been explicitly realized experimentally, essentially in all of its aspects.
But also for nuclear systems the crossover scenario is found to be consistent with various aspects of their phenomenology.
For a review, where the BCS-BEC crossover has been considered on equal footing for ultra-cold Fermi gases and nuclear systems, see Ref.~\cite{Physics-Reports-2018}.
Superconductors are expected to make no exception and to be the next in the list for the relevance of the BCS–BEC crossover.
Actually, the theory of the BCS–BEC crossover took root initially in Ref.~\cite{Eagles-1969}, where possible applications to superconducting semiconductors were envisaged.
Later on, the interest in the BCS–BEC crossover grew up with the advent of high-temperature (cuprate) superconductors, based especially on the argument that the pair size appears to be comparable with the inter-particle spacing.
Nowadays, there is growing evidence for the occurrence of this crossover in condensed-matter systems, like in two-band superconductors with iron-based materials \cite{Kanigel-2012}.
Recently, it was claimed that evidence was collected for the BCS-BEC crossover in the high-temperature superconducting cuprates, by identifying a universal ``magic'' coupling ratio $2 \Delta_{0}/k_{B} T_{c} \approx 6.5$
at which paired fermion condensates become optimally robust \cite{Harrison-Chan-2022}. a
This value should correspond to unitarity in an ultra-cold atomic Fermi gas and strongly deviates from the BCS value $\left[2 \Delta_{0}/k_{B} T_{c}\right]_{\mathrm{BCS}} \approx 3.5$.
As an example that corroborates this argument, \color{red}Fig.~\ref{Figure-6} \color{black} reproduces panel (a) of Fig.~2 from Ref.~\cite{Harrison-Chan-2022}, where the experimental data for the height of the jump $\delta \gamma (T_{c})$ 
in the fermionic (or electronic) contribution $C = \gamma T$ to the specific heat at $T_{c}$ are reported vs the coupling ratio for a number of cuprates.
A theoretical approach that would provide a detailed account for these experimental results is still lacking.

\begin{figure}[t]
\includegraphics[width=9.1cm,angle=0]{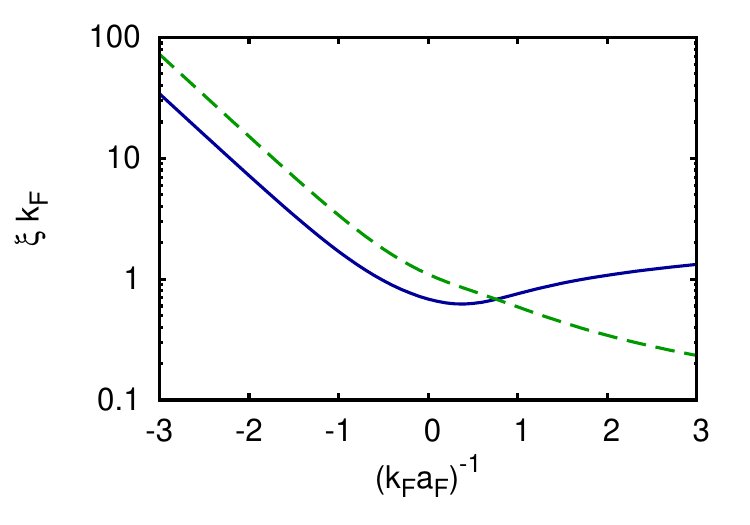}
\caption{Coupling dependence at zero-temperature of the intra-pair coherence length $\xi_{\mathrm{pair}}$ (from Ref.~\cite{PS-1994} - dashed line) and of the inter-pair coherence length $\xi_{\mathrm{phase}}$ (from Ref.~\cite{PS-1996} - full line).
             Both lengths are in units of the average inter-particle distance $k_{F}^{-1}$.
             On the BCS side of unitarity $k_{F} a_{F} \lesssim  0$, the two lengths differ from each other by an irrelevant numerical factor owing to their independent definitions.
             [\emph{Source}: Reproduced from Ref.~\cite{SPS-2010}.]} 
\label{Figure-7}
\end{figure} 

Returning to the argument mentioned above, about the short coherence length (of the order of the inter-particle spacing) which is believed to be associated with high-temperature cuprate superconductors, it should be mentioned that, when spanning the
BCS-BEC crossover from the BCS to the BEC limits, two distinct lengths actually emerge already at zero temperature. 
They are: 
(i) The intra-pair coherence length $\xi_{\mathrm{pair}}$ corresponding to the pair size, which decreases monotonically from the Pippard length $\xi_{0}= k_{F}/(\pi m \Delta_{0})$ in the BCS limit to the bound-state radius $r_{0}=a_{F}/\sqrt{2}$ 
in the BEC limit \cite{PS-1994};
(ii) The inter-pair coherence (or healing) length $\xi_{\mathrm{phase}}$, which coincides with $\xi_{\mathrm{pair}}$ in the BCS limit but accounts for the long-distance coherence among dilute composite bosons in the BEC limit \cite{PS-1996}.
Differences between $\xi_{\mathrm{pair}}$ and $\xi_{\mathrm{phase}}$ become even more marked as a function of temperature at fixed coupling, to the extent that, upon approaching and past $T_{c}$, $\xi_{\mathrm{pair}}$ remains finite
while $\xi_{\mathrm{phase}}$ diverges at $T_{c}$ with a characteristic singular behavior of the superfluid-normal phase transition \cite{PS-2014}.
The coupling dependence of $\xi_{\mathrm{pair}}$ and $\xi_{\mathrm{phase}}$ at zero temperature is shown in \color{red}Fig.~\ref{Figure-7}\color{black}.
Note from this plot that these two lengths coincide with each other in the BCS side and up to unitarity, where they significantly acquire a value comparable with the inter-particle spacing $k_{F}^{-1}$.
Note also that $k_{F} \xi_{\mathrm{pair}}$ is a single-valued function of $(k_{F}a_{F})^{-1}$, such that the two variables can alternatively be used to span the crossover.

\begin{figure}[h]
\includegraphics[width=8.6cm,angle=0]{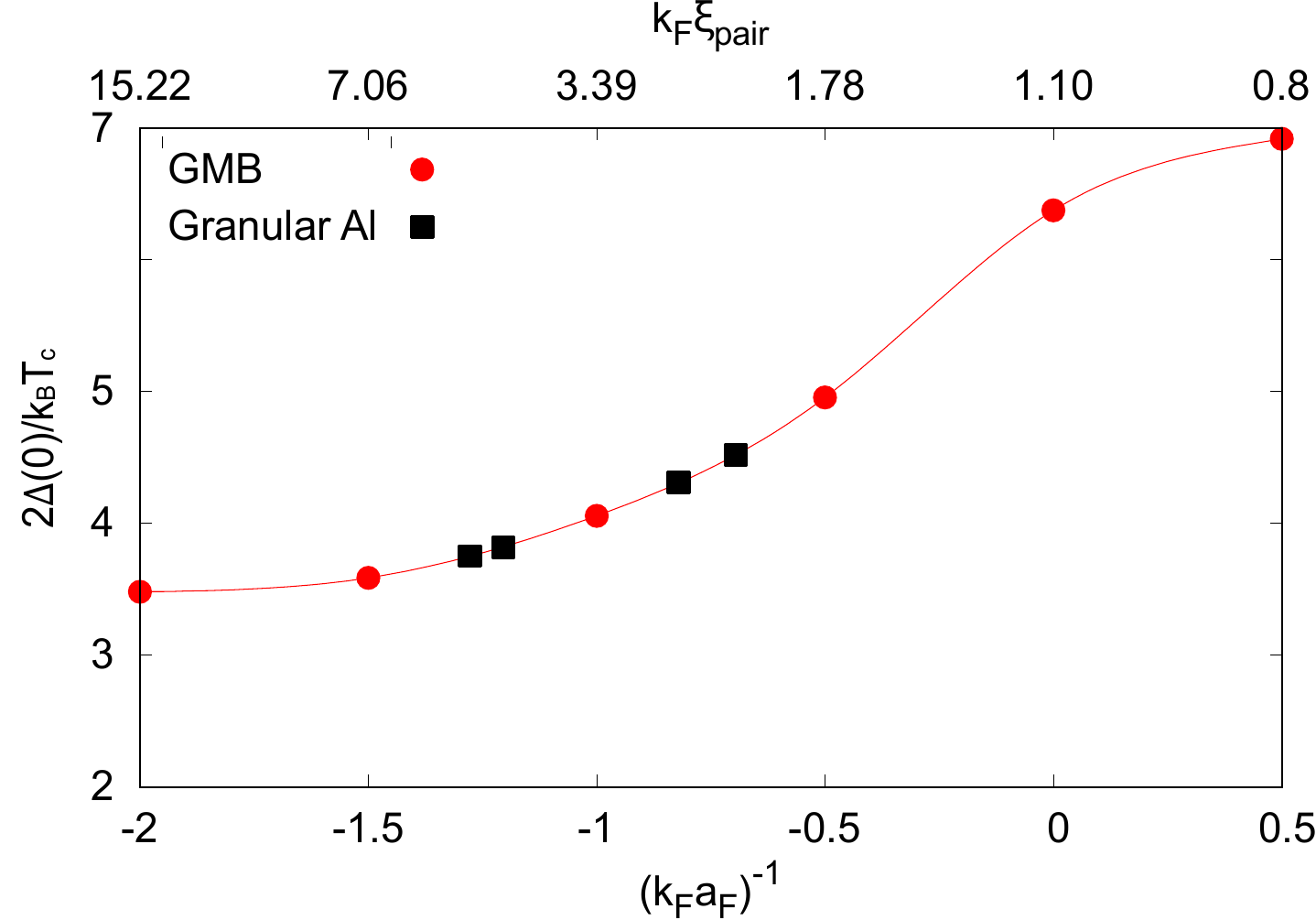}
\caption{The coupling ratio $2 \Delta_{0}/k_{B}T_{c}$ is plotted vs the inter-particle coupling $(k_{F} a_{F})^{-1}$ (lower horizontal axis) and vs $k_{F} \xi_{\mathrm{pair}}$ (upper horizontal axis). 
              The experimental values of this coupling ratio for granular $\mathrm{Al}$ are reported relative to the experimental variable $k_{F} \xi_{\mathrm{pair}}$, while the corresponding theoretical values obtained 
              from the extended GMB approach of Refs.~\cite{PPS-2018} and \cite{PPS-2018} are reported relative to the theoretical variable $(k_{F} a_{F})^{-1}$.
              [\emph{Source}: Reproduced from Ref.~\cite{Deutscher-2019}.]} 
\label{Figure-8}
\end{figure} 

For condensed-matter samples, on the other hand, the inter-particle coupling $(k_{F}a_{F})^{-1}$ cannot be experimentally identified as one does for ultra-cold gases.
In this case, however, it should be possible to describe the crossover in terms of the (zero-temperature) variable $k_{F} \xi_{\mathrm{pair}}$, where $k_{F}$ is obtained in terms of the inter-particle spacing and $\xi_{\mathrm{pair}}$ from upper critical field measurements \cite{Deutscher-2019}.
The dependence of $k_{F} \xi_{\mathrm{pair}}$ on the inter-particle coupling $(k_{F}a_{F})^{-1}$ (as reported in \color{red}Fig.~\ref{Figure-7} \color{black} at low temperature) was nicely exploited in Ref.~\cite{Deutscher-2019}, to relate the measured
values of the coupling ratio $2 \Delta_{0}/k_{B} T_{c}$ to the variable $k_{F} \xi_{\mathrm{pair}}$ and thus to $(k_{F}a_{F})^{-1}$.
These experimental values were further compared with the corresponding theoretical values of $2 \Delta_{0}/k_{B} T_{c}$ vs $(k_{F}a_{F})^{-1}$, obtained from the coupling dependence of $\Delta_{0}$ from Ref.~\cite{PPS-2018} and 
of $T_{c}$ from Ref.~\cite{PPPS-2018}, where they were both obtained by the extended GMB approach.
This comparison is reported in  \color{red}Fig.~\ref{Figure-8}\color{black}.
Even for this case, the comparison between the experimental values and the theoretical results obtained by the extended GMB approach appears rather gratifying.

As a final comment, I would like to mention that, although the present contribution deals with fundamental aspects of strong-coupling fermionic superfluidity/superconductivity aside from what might have produced the strong inter-particle coupling to begin with, and thus it is not concerned with the structural and chemical properties of a given condensed-matter material, I would hope that Prof. Alex M\"{u}ller could have anyway appreciated this contribution as an (albeit partial) attempt to connect the
superfluid properties of physical systems apparently so different from each other (like ultra-cold fermi gases, nuclear systems, and non-conventional superconductors).
In this respect, it should not escape from one's attention that the maximum value $T_{c}/T_{F} \simeq 0.26$ shown above in \color{red}Fig.~\ref{Figure-4} \color{black} is the largest value of the critical temperature attained by a fermionic superfluid.

\vspace{-0.5cm}
\begin{center}
\begin{small}
{\bf ACKNOWLEDGMENTS}
\end{small}
\end{center}

I am indebted to Prof.~G. Deutsher for his long-time interest in the topics considered in the present work through our personal long and fruitful acquaintance.

\newpage
	

\end{document}